\documentclass[aps,prl,a4paper,twocolumn,groupedaddress,showpacs,showkeys,preprintnumbers,floatfix,dvips]{revtex4-1}

\usepackage[utf8]{inputenc} \usepackage[T1]{fontenc}
\usepackage{psfrag} 
\usepackage{times} \usepackage{graphicx} \usepackage{color}
\usepackage{amsmath} \usepackage{amscd} \usepackage{amssymb} \usepackage{amsthm}
\usepackage[section]{placeins} \usepackage[normalem]{ulem}
\usepackage{ragged2e} 
\usepackage{nicefrac}
\usepackage{captcont}
\usepackage{dsfont}
\usepackage{cancel}
\usepackage{calc}  
\usepackage{enumitem}

\begin{document}

\title{Diversity and Disorder in the Voter Model with Delays}
\author{Andr\'e M. Timpanaro} \email[]{a.timpanaro@ufabc.edu.br} \affiliation{Universidade Federal do ABC \\ 09210-580 - Santo Andr\'e -
S\~{a}o Paulo - Brazil} \date{\today}

\pacs{89.65.-s, 02.50.Ey, 02.60.Cb, 05.45.Tb}
\keywords{}

\begin{abstract}
In this work, we investigate interactions that simultaneously order a system locally, while keeping it globally disordered. The study is done in the context of the emergence of diversity in opinion propagation models with interactions rooted in conformity, but some suggestions on how this could be extended to other topics (like ecology and neuroscience) are also made. We do this by introducing a generic modification that can be added to different opinion propagation models (and other agent based models) and that seems to introduce a global tendency towards diversity without leading the system to a frozen state, even in the absence of thermal noises, contrarian agents or cyclic interactions. This modification consists of effectively introducing a relaxation period right after an agent changes its state, during which it cannot change its state again. We tested this modification for the voter model in a square lattice and verified that in the thermodynamic limit, the only attractor is the completely disordered state, where all opinions coexist in the same proportion. For fixed lattice sizes, finite size effects cause a transition with the lenght of the relaxation period between coexistence and consensus. We made simulations for 2, 3, 4, 5 and 6 opinions in a square lattice and mean field calculations for an arbitrary number of opinions.
\end{abstract}

\maketitle

\section{Introduction}

Opinion diversity is an ubiquitous trait of human societies, with the results of democratic voting being the largest scale evidence of them (we can cite the results  from the 2016 US elections and the ensuing demonstrations as a recent example, showing how fragmented opinions can be in a society). In order to reproduce this diversity behaviour, opinion propagation models that revolve around two opinion states (in favour or against an issue) require the introduction of either interactions that behave like a thermal noise (like contrarian agents \cite{Galam-2006} and spontaneous opinion changes \cite{Timpanaro-Fabio-Maycon-Carmen-2015}) with an intensity large enough to be the dominant interaction, or the introduction of a mechanism that freezes the state of the system (like bounded confidence \cite{deffuant-def, HK-def, Schulze-2004, axelrod-def, Timpanaro-2009, Timpanaro-2012} or the CODA mechanism \cite{CODA-GUF}). On the other hand experiments from psychology suggest that interactions that lead to conformity should be the dominant ones, at least in a local level, as this is the most common outcome of discussion in small groups of people \cite{Asch-1951, Moscovici-1969b, Moscovici-Galam-91}.

From a mathematical point of view, the key problem is how to reconcile global disorder with short range interactions that tend to order the system locally, while avoiding reaching a frozen state. This problem is not exclusive to social modelling and also occurs in ecology (while modelling biodiversity \cite{RPS-def}), neuroscience (the neurons in the brain operate in a state in between strong synchronization and lack of synchronization, with deviations from this pattern being responsible for epilepsy and other disorders \cite{epilepsy-detection, epilepsy-periodic}), while also being of potential interest for physical systems (preventing a system from reaching synchronization). To tackle this question we avoided modifications that behave like thermal noises, cyclic interactions between opinions (even though they have natural interpretations in biodiversity models \cite{Tainaka-Itoh-cyclic-voter, RPS-def} they are rather artificial for opinion propagation \cite{Timpanaro-WS-2011}), as well as any restrictions on how opinions interact that create absorbing states other than consensus states.

In order to do so we used two agent states to represent each opinion $X$, that we called susceptible and non-susceptible agents respectively. Susceptible agents with opinion $X$ (denoted by $X_S$) can change their opinions through interaction with other agents, while non-susceptible agents with opinion $X$ (denoted $X_N$) retain their opinion while interacting with other agents.

When an agent changes opinion it adopts the opinion state of the convincing agent, but as a non-susceptible agent ($X_SY\rightarrow Y_NY$). It then can change to a susceptible agent once more, with a given rate, which is equivalent to having a mean relaxation time, or delay, where the agent cannot be convinced again ($X_N \rightarrow X_S$).

This modification can be added to any model that follows the basic structure of one agent convincing another (or copying another agent's state). We decided to study the voter model \cite{votante-def} because its evolution is diffusion-like, preserving magnetisation. This means that the model has no tendency towards neither ordering nor disordering the network. When interpreted as an opinion propagation model this translates to no tendency towards neither a consensus state nor a coexistence state. As such, the voter model allows us to isolate the effect of our modification from other details that more complicated models may have (which is especially true in the mean field calculations). This also means that consensus is reached by a random-walk behaviour, so a modification that prevents the voter model of reaching a consensus state would need to control in some sense how large the resulting random fluctuations are. Adding delays seems to limit the sizes of domains having the same opinion, which can be seen in simulations as the correlation lenght reaching a maximum value.

\section{Model Definition}
As it is usual in the voter model \cite{votante-def}, the society being modelled is represented by a network, where each site represents an agent and the edges represent the possible interactions between them (which is commonly interpreted as which agents know each other). The rules of the voter model after adding delays become:

\begin{itemize}
\item At each time step, choose an agent (site) $i$ at random.
\item With probability $p$, we turn $i$ into a susceptible agent and move to the next time step.
\item With probability $1 - p$, we choose a neighbour $j$ of $i$ at random and $j$ attempts to convince $i$. $j$ is succesful if the agent $i$ is susceptible, in which case $i$ becomes a non-susceptible agent with the same opinion as $j$.
\end{itemize}

The probability $p$ is a parameter of the model, however it is more convenient in the simulations and calculations to control the parameter $\lambda = \nicefrac{p}{(1-p)}$ instead.

We'd like to call attention to the fact that even though we are examining a problem with $M$ opinions, the agents can be in $2M$ different states (susceptible and non-susceptible for each one of the $M$ opinions).

\section{Mean Field Results and Scaling Laws for the Square Lattice}

One can easily write a system of mean field equations from the rules of the voter model with delays and $M$ opinions

\begin{equation}
\left\{
\begin{split}
\dot{\eta}_{\sigma} = -\lambda \eta_{\sigma} + \sum_{{\sigma'}\neq {\sigma}}(\eta_{\sigma} + \nu_{\sigma})\nu_{\sigma'} \\
\dot{\nu}_{\sigma} = \lambda \eta_{\sigma} - \nu_{\sigma} \sum_{{\sigma'}\neq {\sigma}} (\eta_{\sigma'} + \nu_{\sigma'}).
\end{split}
\right.
\label{eq:MF-equations}
\end{equation}
Where $\eta_{\sigma}$ denotes the proportion of agents that have opinion $\sigma$ and are non-susceptibles, while $\nu_{\sigma}$ denotes the proportion of agents that have opinion $\sigma$ and are susceptible.

The fixed points for this system of equations can be easily found. If we define $\Delta$ as the set of surviving opinions ($\sigma \in \Delta \Leftrightarrow \eta_{\sigma} + \nu_{\sigma} \neq 0$) and $\Omega$ as the set of remaining opinions, we have one fixed point for each choice of $\Delta$:

\begin{equation}
\left\{
\begin{split}
&\nu_{\sigma}^{\,*} = \frac{\lambda}{|\Delta|\lambda + |\Delta| - 1}&\mbox{, if }\sigma\in\Delta \\
&\eta_{\sigma}^{\,*} = \frac{|\Delta| - 1}{|\Delta|(|\Delta|\lambda + |\Delta| - 1)}&\mbox{, if }\sigma\in\Delta \\
&\nu_{\sigma}^{\,*} = \eta_{\sigma}^{\,*} = 0&\mbox{, if }\sigma\in\Omega.
\end{split}
\right.
\label{eq:MF-fixed-point}
\end{equation}

A linear stability analysis (done in detail in the Supplementary Material \cite{supp}) reveals that the only attractive fixed point is the one with $\Omega = \varnothing$. This means that the mean-field system evolves towards a situation where all starting opinions are equally represented, independently of the value of $\lambda$ and of the initial conditions.

We now rewrite equation \ref{eq:MF-equations} in order to consider a square lattice geometry (with lattice size scaled to 1):

\begin{equation}
\left\{
\begin{split}
&\dot{\eta}_{\sigma} = - \lambda \eta_{\sigma} + \left( \eta_{\sigma} + \nu_{\sigma} + \frac{L^2}{4}\nabla^2 (\eta_{\sigma} + \nu_{\sigma}) \right) \sum_{\sigma ' \neq \sigma} \nu_{\sigma '} \\
&\dot{\nu}_{\sigma} = \lambda \eta_{\sigma} + \nu_{\sigma} \left(\frac{L^2}{4}\nabla^2 (\eta_{\sigma} + \nu_{\sigma}) -  \sum_{\sigma ' \neq \sigma} (\eta_{\sigma '} + \nu_{\sigma '}) \right)
\end{split}
\right.
\label{eq:forma-normal}
\end{equation}
and use the mean field attractor we just found to cast the equations in a normal form (following \cite{rps-completo}). If

\begin{equation}
\eta^{*}_{\sigma} = \frac{M - 1}{M(\lambda M + M - 1)}\quad\mbox{ and }\quad \nu^{*}_{\sigma} = \frac{\lambda}{\lambda M + M - 1}
\end{equation}
then defining a perturbation $N_{\sigma}$ for $\eta^{*}_{\sigma}$ and $\Theta_{\sigma}$ for $\eta^{*}_{\sigma} + \nu^{*}_{\sigma}$:

\begin{equation}
\eta_{\sigma} = \eta^{*}_{\sigma} + N_{\sigma}\quad\mbox{ and }\quad \eta_{\sigma} + \nu_{\sigma} = \eta^{*}_{\sigma} + \nu^{*}_{\sigma} + \Theta_{\sigma}
\end{equation}
while taking $\lambda \gg 1$ and $N_{\sigma}, \Theta_{\sigma} \ll 1$, leads to (details in \cite{supp})

\begin{equation}
\left\{
\begin{split}
&\dot{N}_{\sigma} = - \lambda N_{\sigma} + \frac{L^2}{4} \nabla^2 \Theta_{\sigma} \\
&\dot{\Theta}_{\sigma} = \lambda N_{\sigma} + \frac{L^2(M-1)}{4M} \nabla^2 \Theta_{\sigma}
\end{split}
\right.
\end{equation}

Now these equations allow us to tell how spatial patterns will scale with $L$ and $\lambda$. Firstly, note that as long as $\nicefrac{\lambda}{L^2}$ remains the same, changing the parameters is equivalent to redefining the time scales, so any steady state properties should be functions of $\nicefrac{\lambda}{L^2}$. Secondly, as we scaled the lattice to have size 1, the sizes of spatial patterns should scale as $L\phi\left(\nicefrac{\lambda}{L^2}\right)$ in the steady state. Moreover, if there are no long range correlations (periodic patterns, for example) we can also argue that these spatial patterns should only depend on $\lambda$ as $L$ grows larger, since they are being generated locally. This allows us to determine $\phi$ and means that the scale should be $\sqrt{\lambda}$. In particular, the correlation lenght $\xi$ should grow with $\sqrt{\lambda}$. This second scaling should be taken with a grain of salt as our deduction assumes $\lambda \gg 1$ and spatial patterns cannot be larger than the lattice itself, so it must be valid only in the regime $1 \ll \lambda \ll L^2$.

Finally, we can estimate the time taken to reach the steady state. Let $\sigma(t)$ be the standard deviation of $\eta_{i}(t) + \nu_{i}(t)$ taken over many simulations with the same initial conditions and after $t$ timesteps. We argue that before the steady state is reached we can approximate our model by a model without delays. Since the usual voter model is equivalent to an unbiased random walk, this leads to $\sigma(t) \simeq \nicefrac{\sqrt{t}}{L^2}$. If a steady state is reached we must have $\sigma(t)$ growing until it reaches some $\sigma_s = \gamma\left(\frac{\lambda}{L^2}\right)$. To determine $\gamma$, we increase the lattice side by a factor $n$, effectively creating an $n\times n$ grid of $L\times L$ lattices. In the steady state, if $L \gg \xi$, each sublattice behaves independently, meaning that $\sigma_s(nL) = \nicefrac{\sigma_s(L)}{n} \Rightarrow \sigma_s \simeq \frac{\sqrt{\lambda}}{L}$. Solving for $\sigma(t_s) = \sigma_s$, yields $t_s \simeq \lambda L^2$ for the amount of timesteps $t_s$ until the steady state is typically reached (which corresponds to an amount of Monte Carlo timesteps dependent on $\lambda$ only).

\section{Simulation Results}

\begin{figure}[htb!]
\begin{center}
\includegraphics[width=\columnwidth]{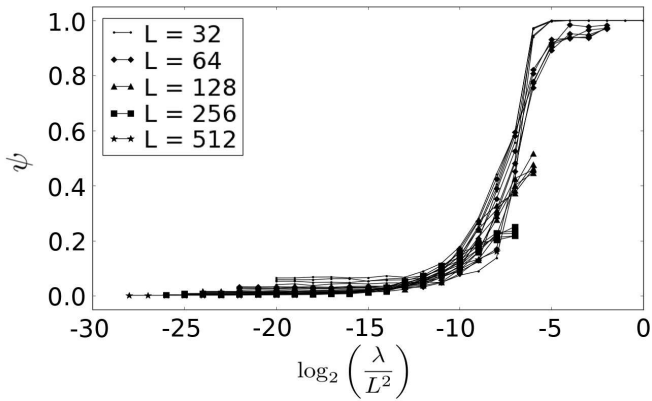}
\caption{Graph of the ensemble average of the order parameter (see eq \ref{eq:psi}) $\psi$ by $\frac{\lambda}{L^2}$ for $M$ ranging from 2 to 6 and $L$ ranging from 32 to 512 and using a fixed amount of monte carlo time steps. We see an approximate collapse of the curves, specially in the region with low $\frac{\lambda}{L^2}$ where the system remains unordered.}
\label{fig:collapse}
\end{center}
\end{figure}

\begin{figure}[htb!]
\begin{center}
\includegraphics[width=\columnwidth]{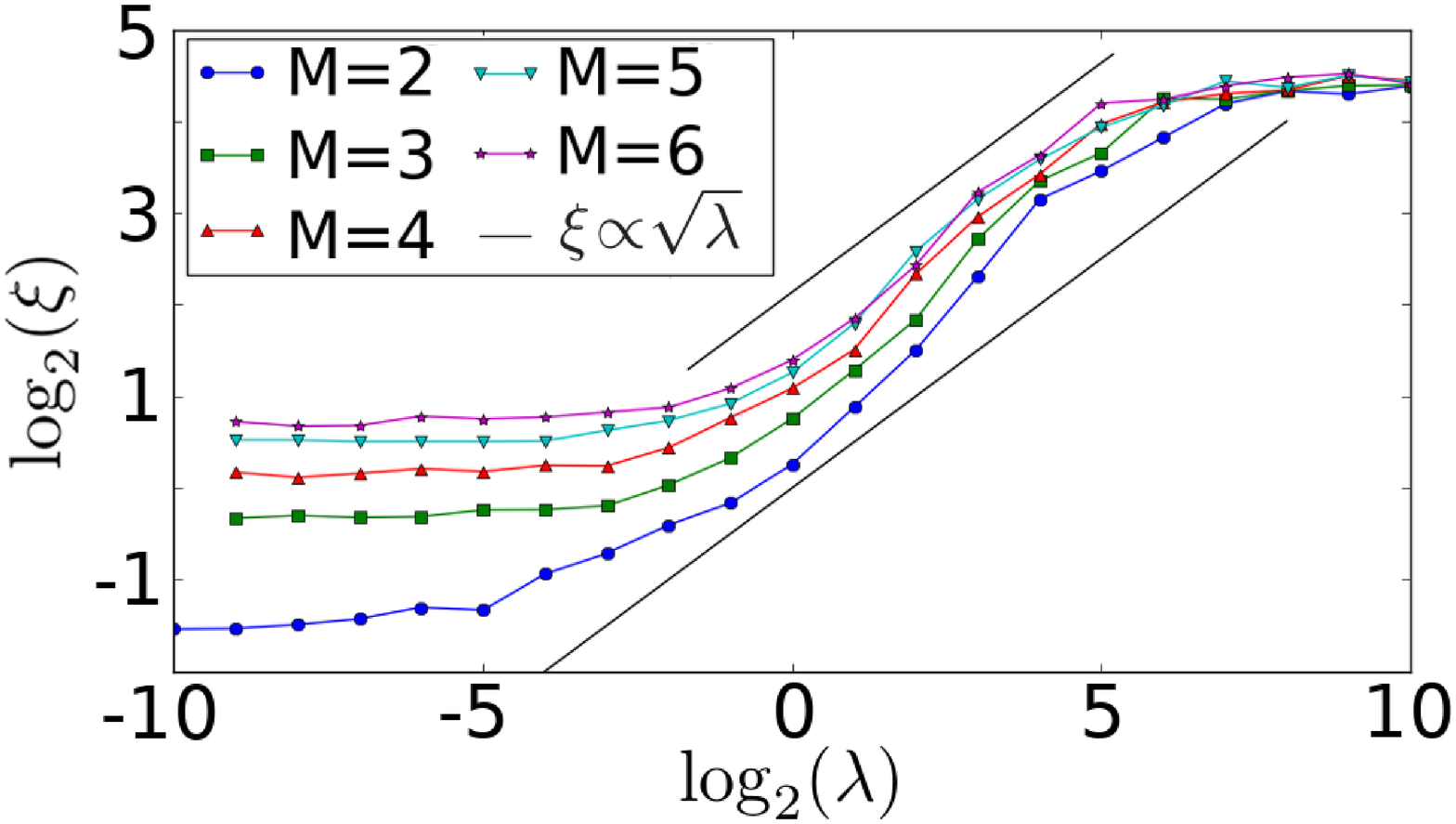}
\caption{Graph of $\log_2(\xi)$ by $\log_2(\lambda)$ for $M$ ranging from 2 to 6 and $L = 256$, together with guiding lines for $\xi \propto \sqrt{\lambda}$. We can see 3 regimes, for medium $\lambda$ we have our prediction $\xi \simeq \sqrt{\lambda}$. For low $\lambda$ the approximation made in our calculations breaks down and $\xi$ deviates from $\sqrt{\lambda}$. For large $\lambda$, the correlation lenght saturates due to the finite size of our lattice.}
\label{fig:corr-p}
\end{center}
\end{figure}

\begin{figure}[htb!]
\begin{center}
\includegraphics[width=\columnwidth]{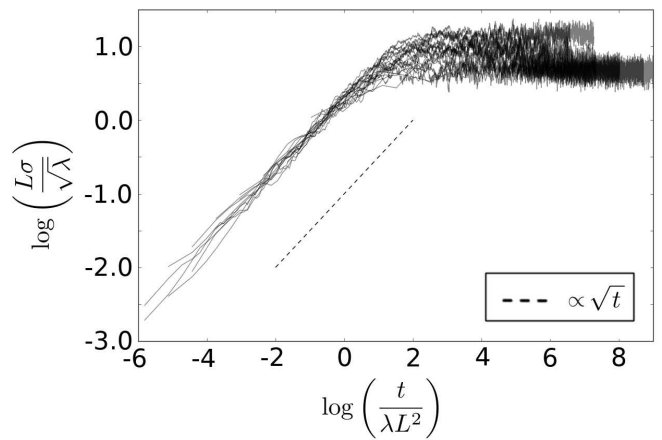}
\caption{Graph of $\log\left(\frac{L\sigma}{\sqrt{\lambda}}\right)$ by $\log\left(\frac{t}{\lambda L^2}\right)$ for $M = 2$, $L$ ranging from 32 to 512 and simulations with $\lambda \leq 256$ (which puts all simulations in the coexistence region of parameter space). Showing that fluctuations reach a maximum value once the steady state is reached.}
\label{fig:sigma-collapse}
\end{center}
\end{figure}

We made simulations of this model in square lattices with sizes ranging from $32\times 32$ to $512\times 512$, for M = 2, \ldots, 6 opinions and the $\lambda$ parameter ranging from $2^{-10}$ to $2^{11}$. Our objective was to use our analytical predictions from the previous section to show the transition between consensus and opinion coexistence. Firstly, we can define the following order parameter for our model:

\begin{equation}
\psi = \frac{M \underset{\sigma}{\max} \{\eta_{\sigma} + \nu_{\sigma}\} - 1}{M - 1}
\label{eq:psi}
\end{equation}
Our predictions imply that if we make the graph of the ensemble average of $\psi$ in the stationary state by $\frac{\lambda}{L^2}$, then the curves will colapse. This can be found in figure \ref{fig:collapse}, where we see that when $\frac{\lambda}{L^2}$ is small the system is unordered. In the same figure we see the predicted colapse, implying a scaling $\lambda_c = \alpha_c L^2$ for the value of $\lambda$ where the transition to consensus happens.

Secondly, we predicted that in the regime $1 \ll \lambda \ll L^2$ the correlation lenght should scale as $\xi \simeq \sqrt{\lambda}$. We looked at the state of the simulation after a steady state was reached and mapped the states in the lattice to 1 if they corresponded to the dominant opinion and 0 otherwise, and used these mapped states in order to obtain the correlation lenght, $\xi$. In figure \ref{fig:corr-p} we have a log-log graph of $\xi$ by $\lambda$, showing 3 regimes. For medium $\lambda$ we have $\xi \simeq \sqrt{\lambda}$ as predicted. For low $\lambda$ the approximation made in our calculations breaks down and $\xi$ deviates from $\sqrt{\lambda}$. For large $\lambda$, the correlation lenght saturates due to the finite size of our lattice.

Finaly, we predicted that the fluctuations grow as $\sigma(t) \simeq \nicefrac{\sqrt{t}}{L^2}$ until $\sigma_s \simeq \frac{\sqrt{\lambda}}{L}$ after a time $t_s \simeq \lambda L^2$, so making a graph of $\nicefrac{L\sigma}{\sqrt{\lambda}}$ by $\nicefrac{t}{\lambda L^2}$ should collapse the curves. This can be seen for $M=2$ in figure \ref{fig:sigma-collapse} (the graphs for the other values of $M$ are similar and can be found in the supplementary material \cite{supp}). This shows that the fluctuations become bounded with the addition of delays.

All these results show that given any number of opinions $M$ and any value of the parameter $\lambda$, in a sufficiently large network we have opinion diversity and hence the consensus simulations found are all due to finite size effects.

\section{Conclusion}

We have found a mechanism for opinion diversity that is fundamentally different from the known ones, such as random opinion changes \cite{Timpanaro-Fabio-Maycon-Carmen-2015}, contrarian agents \cite{Galam-2006}, bounded confidence \cite{deffuant-def, HK-def, Schulze-2004, axelrod-def, Timpanaro-2009, Timpanaro-2012} or cyclic interactions \cite{Timpanaro-WS-2011, Tainaka-Itoh-cyclic-voter, RPS-def}. This mechanism is based on the observation that after someone changes their opinion, there might be a period of time right after the change where they are more resistant to a new opinion change.

This mechanism also avoids some interpretation problems that exist for the usual ones:

\begin{itemize}
\item Unlike with random opinion changes and contrarian agents, it still seems reasonable to consider situations where this mechanism is strong (corresponding to longer delays before someone can change their opinion again) and even in this regime, small groups still reach consensus, as expected from psychology experiments.
\item The observed diversity is structuraly stable, that is adding a small perturbation of the rules doesn't change the qualitative behaviour (this is a consequence of all the fixed points found in our mean field analysis being hyperbolic). This is not the case with bounded confidence, where some opinions are forbidden to interact with each other, but reintroducing a small probability of interaction between them destroys diversity.
\item Finally, the delays are much easier to justify from a social point of view, compared to cyclic interactions.
\end{itemize}

The qualitative behaviour that we saw on the square lattice and in the mean field (corresponding to a complete network) was the same, hinting that the topology of the network may not be very important in this situation. We recall that the voter model has no tendency towards neither a consensus state nor a coexistence state, so that our simulations are showing the isolated effects of this modification. As such, we believe that investigating how it behaves in a model that normally exhibits consensus should be interesting.

As we already stated, this problem of local order competing with global disorder appears in biodiversity and neuroscience as well. We believe that some of our conclusions here are relevant in these contexts too:

\begin{itemize}
\item For ecology models, if we assume that the members of a species are of 3 types: young, adult and old, with spontaneous changes $Y\rightarrow A \rightarrow O$ and adult members being immune to predation, then we arrive at a picture very similar to what we studied. This may allow for models similar to the RPS model \cite{RPS-def} that do away with cyclic interactions but still display biodiversity, which would allow for the simulation of more complex food webs and reduce the dependance of the diversity with the spatial patterns created in the lattice.

\item For neuroscience, our results may be an indication that the relaxation time that neurons exhibit in between firings \cite{Rulkov-def, FitzHugh-Nagumo-def} may play a role in preventing the brain from entering an epileptic state (where synchronization between neurons is stronger than normal \cite{epilepsy-detection, epilepsy-periodic}).
\end{itemize}

\bibliographystyle{apsrev4-1}
\bibliography{andre}

\begin{thebibliography}{22}%
\makeatletter
\providecommand \@ifxundefined [1]{%
 \@ifx{#1\undefined}
}%
\providecommand \@ifnum [1]{%
 \ifnum #1\expandafter \@firstoftwo
 \else \expandafter \@secondoftwo
 \fi
}%
\providecommand \@ifx [1]{%
 \ifx #1\expandafter \@firstoftwo
 \else \expandafter \@secondoftwo
 \fi
}%
\providecommand \natexlab [1]{#1}%
\providecommand \enquote  [1]{``#1''}%
\providecommand \bibnamefont  [1]{#1}%
\providecommand \bibfnamefont [1]{#1}%
\providecommand \citenamefont [1]{#1}%
\providecommand \href@noop [0]{\@secondoftwo}%
\providecommand \href [0]{\begingroup \@sanitize@url \@href}%
\providecommand \@href[1]{\@@startlink{#1}\@@href}%
\providecommand \@@href[1]{\endgroup#1\@@endlink}%
\providecommand \@sanitize@url [0]{\catcode `\\12\catcode `\$12\catcode
  `\&12\catcode `\#12\catcode `\^12\catcode `\_12\catcode `\%12\relax}%
\providecommand \@@startlink[1]{}%
\providecommand \@@endlink[0]{}%
\providecommand \url  [0]{\begingroup\@sanitize@url \@url }%
\providecommand \@url [1]{\endgroup\@href {#1}{\urlprefix }}%
\providecommand \urlprefix  [0]{URL }%
\providecommand \Eprint [0]{\href }%
\providecommand \doibase [0]{http://dx.doi.org/}%
\providecommand \selectlanguage [0]{\@gobble}%
\providecommand \bibinfo  [0]{\@secondoftwo}%
\providecommand \bibfield  [0]{\@secondoftwo}%
\providecommand \translation [1]{[#1]}%
\providecommand \BibitemOpen [0]{}%
\providecommand \bibitemStop [0]{}%
\providecommand \bibitemNoStop [0]{.\EOS\space}%
\providecommand \EOS [0]{\spacefactor3000\relax}%
\providecommand \BibitemShut  [1]{\csname bibitem#1\endcsname}%
\let\auto@bib@innerbib\@empty
%</preamble>
\bibitem [{\citenamefont {Borghesi}\ and\ \citenamefont
  {Galam}(2006)}]{Galam-2006}%
  \BibitemOpen
  \bibfield  {author} {\bibinfo {author} {\bibfnamefont {C.}~\bibnamefont
  {Borghesi}}\ and\ \bibinfo {author} {\bibfnamefont {S.}~\bibnamefont
  {Galam}},\ }\href@noop {} {\bibfield  {journal} {\bibinfo  {journal}
  {Physical Review E}\ }\textbf {\bibinfo {volume} {73}},\ \bibinfo {pages}
  {066118} (\bibinfo {year} {2006})}\BibitemShut {NoStop}%
\bibitem [{\citenamefont {Ara\'ujo}\ \emph {et~al.}(2015)\citenamefont
  {Ara\'ujo}, \citenamefont {Vannucchi}, \citenamefont {Timpanaro},\ and\
  \citenamefont {Prado}}]{Timpanaro-Fabio-Maycon-Carmen-2015}%
  \BibitemOpen
  \bibfield  {author} {\bibinfo {author} {\bibfnamefont {M.~S.}\ \bibnamefont
  {Ara\'ujo}}, \bibinfo {author} {\bibfnamefont {F.~S.}\ \bibnamefont
  {Vannucchi}}, \bibinfo {author} {\bibfnamefont {A.~M.}\ \bibnamefont
  {Timpanaro}}, \ and\ \bibinfo {author} {\bibfnamefont {C.~P.~C.}\
  \bibnamefont {Prado}},\ }\href {\doibase 10.1103/PhysRevE.91.022813}
  {\bibfield  {journal} {\bibinfo  {journal} {Phys. Rev. E}\ }\textbf {\bibinfo
  {volume} {91}},\ \bibinfo {pages} {022813} (\bibinfo {year}
  {2015})}\BibitemShut {NoStop}%
\bibitem [{\citenamefont {Deffuant}\ \emph {et~al.}(2000)\citenamefont
  {Deffuant}, \citenamefont {Neau}, \citenamefont {Amblard},\ and\
  \citenamefont {Weisbuch}}]{deffuant-def}%
  \BibitemOpen
  \bibfield  {author} {\bibinfo {author} {\bibfnamefont {G.}~\bibnamefont
  {Deffuant}}, \bibinfo {author} {\bibfnamefont {D.}~\bibnamefont {Neau}},
  \bibinfo {author} {\bibfnamefont {F.}~\bibnamefont {Amblard}}, \ and\
  \bibinfo {author} {\bibfnamefont {G.}~\bibnamefont {Weisbuch}},\ }\href@noop
  {} {\bibfield  {journal} {\bibinfo  {journal} {Advances in Complex Systems}\
  }\textbf {\bibinfo {volume} {3}},\ \bibinfo {pages} {87} (\bibinfo {year}
  {2000})}\BibitemShut {NoStop}%
\bibitem [{\citenamefont {Hegselmann}\ and\ \citenamefont
  {Krause}(2002)}]{HK-def}%
  \BibitemOpen
  \bibfield  {author} {\bibinfo {author} {\bibfnamefont {R.}~\bibnamefont
  {Hegselmann}}\ and\ \bibinfo {author} {\bibfnamefont {U.}~\bibnamefont
  {Krause}},\ }\href@noop {} {\bibfield  {journal} {\bibinfo  {journal}
  {Journal of Artificial Societies and Social Simulation}\ }\textbf {\bibinfo
  {volume} {5}} (\bibinfo {year} {2002})}\BibitemShut {NoStop}%
\bibitem [{\citenamefont {Schulze}(2004)}]{Schulze-2004}%
  \BibitemOpen
  \bibfield  {author} {\bibinfo {author} {\bibfnamefont {C.}~\bibnamefont
  {Schulze}},\ }\href@noop {} {\bibfield  {journal} {\bibinfo  {journal}
  {International Journal of Modern Physics C}\ }\textbf {\bibinfo {volume}
  {15}},\ \bibinfo {pages} {867} (\bibinfo {year} {2004})}\BibitemShut
  {NoStop}%
\bibitem [{\citenamefont {Axelrod}(1997)}]{axelrod-def}%
  \BibitemOpen
  \bibfield  {author} {\bibinfo {author} {\bibfnamefont {R.}~\bibnamefont
  {Axelrod}},\ }\href@noop {} {\bibfield  {journal} {\bibinfo  {journal} {The
  Journal of Conflict Resolution}\ }\textbf {\bibinfo {volume} {41}},\ \bibinfo
  {pages} {203} (\bibinfo {year} {1997})}\BibitemShut {NoStop}%
\bibitem [{\citenamefont {Timpanaro}\ and\ \citenamefont
  {do~Prado}(2009)}]{Timpanaro-2009}%
  \BibitemOpen
  \bibfield  {author} {\bibinfo {author} {\bibfnamefont {A.~M.}\ \bibnamefont
  {Timpanaro}}\ and\ \bibinfo {author} {\bibfnamefont {C.~P.~C.}\ \bibnamefont
  {do~Prado}},\ }\href@noop {} {\bibfield  {journal} {\bibinfo  {journal}
  {Physical Review E}\ }\textbf {\bibinfo {volume} {80}},\ \bibinfo {pages}
  {021119} (\bibinfo {year} {2009})}\BibitemShut {NoStop}%
\bibitem [{\citenamefont {Timpanaro}\ and\ \citenamefont
  {do~Prado}(2012)}]{Timpanaro-2012}%
  \BibitemOpen
  \bibfield  {author} {\bibinfo {author} {\bibfnamefont {A.~M.}\ \bibnamefont
  {Timpanaro}}\ and\ \bibinfo {author} {\bibfnamefont {C.~P.~C.}\ \bibnamefont
  {do~Prado}},\ }\href@noop {} {\bibfield  {journal} {\bibinfo  {journal}
  {Physical Review E}\ }\textbf {\bibinfo {volume} {86}},\ \bibinfo {pages}
  {046109} (\bibinfo {year} {2012})}\BibitemShut {NoStop}%
\bibitem [{\citenamefont {Martins}\ and\ \citenamefont
  {Galam}(2013)}]{CODA-GUF}%
  \BibitemOpen
  \bibfield  {author} {\bibinfo {author} {\bibfnamefont {A.~C.~R.}\
  \bibnamefont {Martins}}\ and\ \bibinfo {author} {\bibfnamefont
  {S.}~\bibnamefont {Galam}},\ }\href {\doibase 10.1103/PhysRevE.87.042807}
  {\bibfield  {journal} {\bibinfo  {journal} {Phys. Rev. E}\ }\textbf {\bibinfo
  {volume} {87}},\ \bibinfo {pages} {042807} (\bibinfo {year}
  {2013})}\BibitemShut {NoStop}%
\bibitem [{\citenamefont {Asch}(1951)}]{Asch-1951}%
  \BibitemOpen
  \bibfield  {author} {\bibinfo {author} {\bibfnamefont {S.~E.}\ \bibnamefont
  {Asch}},\ }in\ \href@noop {} {\emph {\bibinfo {booktitle} {Groups, leadership
  and men.}}},\ \bibinfo {editor} {edited by\ \bibinfo {editor} {\bibfnamefont
  {H.}~\bibnamefont {Guetzkow}}}\ (\bibinfo  {publisher} {Carnegie Press},\
  \bibinfo {address} {Pittsburgh, PA},\ \bibinfo {year} {1951})\BibitemShut
  {NoStop}%
\bibitem [{\citenamefont {Moscovici}\ and\ \citenamefont
  {Zavalloni}(1969)}]{Moscovici-1969b}%
  \BibitemOpen
  \bibfield  {author} {\bibinfo {author} {\bibfnamefont {S.}~\bibnamefont
  {Moscovici}}\ and\ \bibinfo {author} {\bibfnamefont {M.}~\bibnamefont
  {Zavalloni}},\ }\href@noop {} {\bibfield  {journal} {\bibinfo  {journal}
  {Journal of Personality and Social Psychology}\ }\textbf {\bibinfo {volume}
  {12}},\ \bibinfo {pages} {125} (\bibinfo {year} {1969})}\BibitemShut
  {NoStop}%
\bibitem [{\citenamefont {Galam}\ and\ \citenamefont
  {Moscovici}(1991)}]{Moscovici-Galam-91}%
  \BibitemOpen
  \bibfield  {author} {\bibinfo {author} {\bibfnamefont {S.}~\bibnamefont
  {Galam}}\ and\ \bibinfo {author} {\bibfnamefont {S.}~\bibnamefont
  {Moscovici}},\ }\href {\doibase 10.1002/ejsp.2420210105} {\bibfield
  {journal} {\bibinfo  {journal} {European Journal of Social Psychology}\
  }\textbf {\bibinfo {volume} {21}},\ \bibinfo {pages} {49} (\bibinfo {year}
  {1991})}\BibitemShut {NoStop}%
\bibitem [{\citenamefont {Reichenbach}\ \emph {et~al.}(2007)\citenamefont
  {Reichenbach}, \citenamefont {Mobilia},\ and\ \citenamefont
  {Frey}}]{RPS-def}%
  \BibitemOpen
  \bibfield  {author} {\bibinfo {author} {\bibfnamefont {T.}~\bibnamefont
  {Reichenbach}}, \bibinfo {author} {\bibfnamefont {M.}~\bibnamefont
  {Mobilia}}, \ and\ \bibinfo {author} {\bibfnamefont {E.}~\bibnamefont
  {Frey}},\ }\href@noop {} {\bibfield  {journal} {\bibinfo  {journal} {Nature}\
  }\textbf {\bibinfo {volume} {448}},\ \bibinfo {pages} {1046} (\bibinfo {year}
  {2007})}\BibitemShut {NoStop}%
\bibitem [{\citenamefont {Jerger}\ \emph {et~al.}(2005)\citenamefont {Jerger},
  \citenamefont {Weinstein}, \citenamefont {Sauer},\ and\ \citenamefont
  {Schiff}}]{epilepsy-detection}%
  \BibitemOpen
  \bibfield  {author} {\bibinfo {author} {\bibfnamefont {K.~K.}\ \bibnamefont
  {Jerger}}, \bibinfo {author} {\bibfnamefont {S.~L.}\ \bibnamefont
  {Weinstein}}, \bibinfo {author} {\bibfnamefont {T.}~\bibnamefont {Sauer}}, \
  and\ \bibinfo {author} {\bibfnamefont {S.~J.}\ \bibnamefont {Schiff}},\
  }\href {\doibase https://doi.org/10.1016/j.clinph.2004.08.023} {\bibfield
  {journal} {\bibinfo  {journal} {Clinical Neurophysiology}\ }\textbf {\bibinfo
  {volume} {116}},\ \bibinfo {pages} {545 } (\bibinfo {year}
  {2005})}\BibitemShut {NoStop}%
\bibitem [{\citenamefont {Penfield}\ and\ \citenamefont
  {Jasper}(1954)}]{epilepsy-periodic}%
  \BibitemOpen
  \bibfield  {author} {\bibinfo {author} {\bibfnamefont {W.}~\bibnamefont
  {Penfield}}\ and\ \bibinfo {author} {\bibfnamefont {H.}~\bibnamefont
  {Jasper}},\ }\href@noop {} {\emph {\bibinfo {title} {Epilepsy and the
  Functional Anatomy of the Human Brain}}}\ (\bibinfo  {publisher} {Little,
  Brown},\ \bibinfo {address} {Brown},\ \bibinfo {year} {1954})\BibitemShut
  {NoStop}%
\bibitem [{\citenamefont {Tainaka}\ and\ \citenamefont
  {Itoh}(1991)}]{Tainaka-Itoh-cyclic-voter}%
  \BibitemOpen
  \bibfield  {author} {\bibinfo {author} {\bibfnamefont {K.}~\bibnamefont
  {Tainaka}}\ and\ \bibinfo {author} {\bibfnamefont {Y.}~\bibnamefont {Itoh}},\
  }\href@noop {} {\bibfield  {journal} {\bibinfo  {journal} {Europhysics
  Letters}\ }\textbf {\bibinfo {volume} {15}},\ \bibinfo {pages} {399}
  (\bibinfo {year} {1991})}\BibitemShut {NoStop}%
\bibitem [{\citenamefont {Timpanaro}\ and\ \citenamefont
  {do~Prado}(2011)}]{Timpanaro-WS-2011}%
  \BibitemOpen
  \bibfield  {author} {\bibinfo {author} {\bibfnamefont {A.~M.}\ \bibnamefont
  {Timpanaro}}\ and\ \bibinfo {author} {\bibfnamefont {C.~P.~C.}\ \bibnamefont
  {do~Prado}},\ }\href@noop {} {\bibfield  {journal} {\bibinfo  {journal}
  {Physical Review E}\ }\textbf {\bibinfo {volume} {84}},\ \bibinfo {pages}
  {027101} (\bibinfo {year} {2011})}\BibitemShut {NoStop}%
\bibitem [{\citenamefont {Holley}\ and\ \citenamefont
  {Liggett}(1975)}]{votante-def}%
  \BibitemOpen
  \bibfield  {author} {\bibinfo {author} {\bibfnamefont {R.~A.}\ \bibnamefont
  {Holley}}\ and\ \bibinfo {author} {\bibfnamefont {T.~M.}\ \bibnamefont
  {Liggett}},\ }\href@noop {} {\bibfield  {journal} {\bibinfo  {journal}
  {Annals of Probability}\ }\textbf {\bibinfo {volume} {3}},\ \bibinfo {pages}
  {643} (\bibinfo {year} {1975})}\BibitemShut {NoStop}%
\bibitem [{\citenamefont {{See Supplemental Material at [...] for the remaining
  graphs and more detailed calculations}}()}]{supp}%
  \BibitemOpen
  \bibfield  {author} {\bibinfo {author} {\bibnamefont {{See Supplemental
  Material at [...] for the remaining graphs and more detailed
  calculations}}},\ }\href@noop {} {}\BibitemShut {NoStop}%
\bibitem [{\citenamefont {Reichenbach}\ \emph {et~al.}(2008)\citenamefont
  {Reichenbach}, \citenamefont {Mobilia},\ and\ \citenamefont
  {Frey}}]{rps-completo}%
  \BibitemOpen
  \bibfield  {author} {\bibinfo {author} {\bibfnamefont {T.}~\bibnamefont
  {Reichenbach}}, \bibinfo {author} {\bibfnamefont {M.}~\bibnamefont
  {Mobilia}}, \ and\ \bibinfo {author} {\bibfnamefont {E.}~\bibnamefont
  {Frey}},\ }\href@noop {} {\bibfield  {journal} {\bibinfo  {journal} {Journal
  of Theoretical Biology}\ }\textbf {\bibinfo {volume} {254}},\ \bibinfo
  {pages} {368} (\bibinfo {year} {2008})}\BibitemShut {NoStop}%
\bibitem [{\citenamefont {Rulkov}(2002)}]{Rulkov-def}%
  \BibitemOpen
  \bibfield  {author} {\bibinfo {author} {\bibfnamefont {N.~F.}\ \bibnamefont
  {Rulkov}},\ }\href {\doibase 10.1103/PhysRevE.65.041922} {\bibfield
  {journal} {\bibinfo  {journal} {Phys. Rev. E}\ }\textbf {\bibinfo {volume}
  {65}},\ \bibinfo {pages} {041922} (\bibinfo {year} {2002})}\BibitemShut
  {NoStop}%
\bibitem [{\citenamefont {FitzHugh}(1955)}]{FitzHugh-Nagumo-def}%
  \BibitemOpen
  \bibfield  {author} {\bibinfo {author} {\bibfnamefont {R.}~\bibnamefont
  {FitzHugh}},\ }\href {\doibase 10.1007/BF02477753} {\bibfield  {journal}
  {\bibinfo  {journal} {The bulletin of mathematical biophysics}\ }\textbf
  {\bibinfo {volume} {17}},\ \bibinfo {pages} {257} (\bibinfo {year}
  {1955})}\BibitemShut {NoStop}%
\end{thebibliography}%

\end{document}